\documentclass[prd,aps,preprintnumbers,twocolumn,showpacs, nofootinbib,superscriptaddress,notitlepage]{revtex4-1}
\usepackage[normalem]{ulem}
\usepackage{epsfig}
\usepackage{amsfonts}
\usepackage{amsmath}
\usepackage{slashed}
\usepackage{graphicx}
\usepackage{color}
\usepackage{hyperref} 
\usepackage{mathtools}
\allowdisplaybreaks[4]

\newcommand{\vect}{\boldsymbol}

\begin{document}
	
	
	\title{Production of true para-muonium in linearly polarized photon fusions}

	\author{Jian-Ping Dai}
    \email{daijianping@ynu.edu.cn}
\affiliation{Department of Physics, Yunnan University, Kunming 650091, China}
		
	\author{Shuai Zhao}
	\email{Corresponding author: zhaos@tju.edu.cn}
\affiliation{Department of Physics, Tianjin University, Tianjin 300350, China}


\begin{abstract}

True muonium (TM)---the bound state of $\mu^+\mu^-$---has not been discovered yet. It was demonstrated that searching for TM via $\gamma\gamma$ fusions in heavy ion collisions is feasible due to the enhancement of the atom number. We study the production of the true para-muonium (pTM) in the collisions of linearly polarized photons in the experiments of heavy-ion collisions, calculate the production rate as well as the transverse spectrum of pTM, and explore the discovery potential in nuclear experiments. Our results show that there is a significant correlation between 
the linearly polarized photon distribution and the transverse momentum distribution of pTM. 
The optimal kinematic region of the generated pTM is identified, which can provide a theoretical guide to the detection of pTM in experiments.
\end{abstract}

	\maketitle

The $\mu^+\mu^-$ bound state, referred to as true muonium (TM), is a short-lived, exotic atomlike particle composed of a muon and its antiparticle, the antimuon. It is a purely quantum electrodynamics (QED) bound state. 
The lightest QED atom, the positronium ($e^+e^-$ bound state), was observed more than 75 years ago~\cite{Deutsch:1951zza} and studied extensively. Even the so-called muonium ($\mu e$ bound state)~\cite{Hughes:1960zz}, $\pi\mu$ atom~\cite{Coombes:1976hi}, and the
dipositronium~\cite{Cassidy:2007} [$(e^+e^-)(e^+e^-)$ molecule] have been discovered and studied more, while the  TM remains unobserved, even though the possibility of its existence was discussed~\cite{Marshak:1947zz} soon after the clarification of the leptonic nature of the muon~\cite{Lattes:1947mw,Lattes:1947mx,Lattes:1947my} in the mid-20th century. The investigation of TM can contribute to our broader understanding of the Standard Model of particle physics and may offer clues to the existence of physics beyond this well-established framework~\cite{Jaeckel:2010xx,Tucker-Smith:2010wdq,Kopp:2014tsa,Agrawal:2021dbo}.

Muons are heavy, unstable relatives of electrons, with a mass approximately 207 times that of an electron. Due to the short lifetime of muons and their antimatter counterparts, TM exists only fleetingly~\cite{Bilenkii:1969,Hughes:1971,Malenfant:1987tm,Jentschura:1998vkm},
making it challenging to study and observe directly. The muon's relatively short lifetime, on the order of microseconds~\cite{ParticleDataGroup:2022pth}, restricts the timescale for the formation and stability of TM. Many sophisticated experimental techniques have been employed to detect
and study these ephemeral particles, often relying on high-energy accelerators and advanced detection methods. It was proposed to detect the TM in the physical processes at high-energy experimental apparatus, such as $\pi^- p\to (\mu^+ \mu^-)n$~\cite{Bilenkii:1969}, $\gamma Z\to (\mu^+ \mu^-)Z$~\cite{Bilenkii:1969,Francener:2021wzx}, $e Z\to e(\mu^+ \mu^-)Z$~\cite{Arteaga-Romero:2000mwd,Krachkov:2017afm,Banburski:2012tk}, $Z_1 Z_2\to Z_1 Z_2(\mu^+\mu^-)$~\cite{Ginzburg:1998df,Chen:2012ci,Yu:2013uka,Azevedo:2019hqp,Yu:2022hdt,Francener:2021wzx} (where $Z$ indicates a heavy nucleus), $\mu^+\mu^-$ collisions~\cite{Hughes:1971}, $\eta\to (\mu^+\mu^-)\gamma$~\cite{Nemenov:1972,CidVidal:2019qub}, $e^+ e^-$ collisions~\cite{Moffat:1975uw,Brodsky:2009gx,Gargiulo:2023tci}, $K_L\to (\mu^+\mu^-)\gamma$~\cite{Ji:2017lyh}, etc. Among these processes, the $e^+ e^-\to (\mu^+ \mu^-)$ and $\gamma (\mu^+\mu^-)$ is of particular interest because there is no hadron involved and thus is a pure QED process. It was demonstrated in Ref.~\cite{Brodsky:2009gx} that the TM could be discovered at electron-positron colliders, if the $\mu^+\mu^-$
resonances above the threshold are also taken into account, because the states with high-principal quantum numbers cannot be distinguished from the resonances just above the threshold due to the beam energy spread, which production rates are enhanced by the Sommerfeld-Schwinger-Sakharov factor~\cite{Sommerfeld:1939}. Based on this, it was proposed to search for TM with fool's intersection storage rings [$e^+e^-\to (\mu^+\mu^-)$] discussed by Bjorken~\cite{Bjorken:1976mk} in which the electron and positron beams are arranged to merge at a small angle in order to give rise to a strong boost for the produced TM, and to search for the true para-muonium (pTM) and true ortho-muonium (oTM) using the initial and final radiation technology [$e^+e^-\to \gamma (\mu^+\mu^-)$] on BESIII and Belle-II experiments~\cite{Harris:2008tx,Belle-II:2010dht}. 

Another promising process is the heavy-ion collisions. If the heavy ions collide with impact parameter larger than the twice of the radius of the nuclei, i.e., the ultraperipheral collisions (UPC), the strong interaction is suppressed, ensuring $\gamma\gamma$ fusion in UPC a QED dominant process~\cite{Baltz:2007kq}. The electromagnetic field strength is scaled with the atom number $Z$. Hence comparing with electron-positron collisions, the production rate is enhanced by a factor $Z^4$ in $\gamma\gamma$ fusions in UPC. For example, in lead-lead collisions the rate will be enhanced by $82^4\sim 4.5\times 10^7$. This makes the study of rare processes possible in heavy-ion collisions. As a consequence, the light-by-light scattering, predicted by quantum theory but forbidden by classical electrodynamics, was observed in $\mathrm{Pb}+\mathrm{Pb}(\gamma\gamma)\to \mathrm{Pb}^*+\mathrm{Pb}^*+\gamma\gamma$ process by ATLAS~\cite{ATLAS:2017fur} and CMS~\cite{CMS:2018erd}, based on the approach proposed in~\cite{dEnterria:2013zqi}. 

On the theory side, the calculation can be carried out with the equivalent photon approximation where the virtual photons are regarded as quasi-real photons, which is also known as the Weizs\"{a}cker--Williams method~\cite{Fermi:1924tc,vonWeizsacker:1934nji,Williams:1935dka}.
The quasi-real photons, radiated from the relativistic nuclei, can be either unpolarized and linearly polarized. The linearly polarized photon distribution is correlated with the transverse momentum and the azimuthal distribution of the produced particles. It was proposed that the linearly polarized photon distribution can be extracted from the azimuthal-asymmetries of the lepton pair production in UPCs, see, e.g., Ref~\cite{Li:2019yzy}. The $\cos4\phi$ asymmetry has been measured by STAR collaboration~\cite{STAR:2019wlg}.
 
The transverse momentum distribution provides useful information for experimental research. Although the total production rate of pTM in heavy-ion collisions has been studied by previous works,
however, the transverse momentum distribution of pTM has never been extensively explored, especially in the small transverse momentum region. Moreover, the effect of the linearly polarized photon is never explored. In this work, We will calculate the transverse momentum dependent differential cross-sections and study the effect of linearly polarized photons on the transverse momentum distribution of the pTM.

\begin{figure}[!htbp]
	\begin{center}
		\includegraphics[width=0.9 \linewidth]{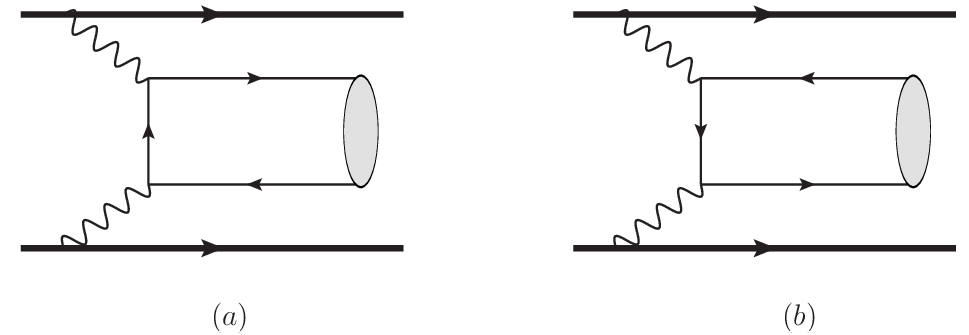}
	\end{center}
	\caption{Production of true para-muonium via $\gamma\gamma$ fusion in heavy-ion collisions.}\label{fig:gammagamma}
\end{figure}

The pTM---the $^1S_0$ bound state of $\mu^-\mu^+$, can be produced by $\gamma\gamma$ fusion, which is shown in Fig.~\ref{fig:gammagamma}.
When the virtuality of photon is small, one can utilize the equivalent photon approximation to calculate the differential cross section. In this work, we take the transverse momentum distribution of pTM into account. If the transverse momentum of pTM in the final state is small, the differential cross section takes the form of transverse-momentum-dependent (TMD) factorization~\cite{Bacchetta:2007sz,Boer:2011kf,Boer:2012bt},
\begin{align}
	d\sigma=&\frac{1}{2s}\frac{d^3 \vect{q}}{(2\pi)^3 2q^0} \int dx_a dx_b d^2 \vect{p}_{aT}d^2 \vect{p}_{bT} \nonumber\\
	&(2\pi)^4 \delta^4 (p_a+p_b-q) \Gamma^{\mu\nu}_A (x_a, \vect{p}_{aT})\Gamma^{\rho\sigma}_B(x_b, \vect{p}_{bT}) \nonumber\\
	&\times H_{\mu\rho;\nu\sigma}(p_a,p_b;q), 
\end{align}
where $\sqrt{s}$ is the collision energy per nucleon pair in the center-of-mass frame,  {$p_a$ and $p_b$ are the momenta of the photons from nucleus $A$ and $B$, respectively, and $q$ is the momentum of pTM.} $H$ is the partonic scattering amplitude of $\gamma\gamma\to \rm{pTM}$ multiplied by its complex conjugate, and $\Gamma_{A,B}$ is the photon correlation matrix of the nuclei $A$ and $B$, respectively, defined as the Fourier transform of the correlator of electromagnetic tensor as
\begin{align}
	&\Gamma^{\mu\nu}(x,\vect{k}_T)\nonumber\\
	=&	\int\frac{d z^- dz_{T}^2}{x^2 P^+ (2\pi)^3} e^{i k \cdot z}\langle P| F^{+\mu}(0) F^{+\nu}(z)|P\rangle |_{z^+=0}\nonumber\\
	=&-\frac{1}{2x} g_{T}^{\mu\nu} f_1^{\gamma} (x, \vect{k}_{T}^2)+\frac{1}{x}\left(\frac12 g^{\mu\nu}_{T}+\frac{k_{T}^{\mu}k_{T}^{\nu}}{\vect{k}_{T}^2}\right) h_1^{\perp \gamma} (x, \vect{k}_{T}^2),
\end{align}
where $g_T^{\mu\nu}=g^{\mu\nu}-n^{\mu}\bar n^{\nu}-n^{\nu}\bar n^{\mu}$ is the transverse metric and the light cone vectors are $n=\frac{1}{\sqrt{2}}(1,0,0,-1)$ and $\bar n=\frac{1}{\sqrt{2}}(1,0,0,1)$. The transverse part of a vector $a$ is then expressed as $a_T^{\mu}=g_T^{\mu\nu}a_{\nu}$.
$F^{\mu\nu}=\partial^{\mu}A^{\nu}-\partial^{\nu}A^{\mu}$ is the electromagnetic tensor, and $z=z^- \bar n+z_T$; distinguished from the gluon case,  there is no Wilson line.  $f_1^{\gamma}$ is the unpolarized photon distribution function, while $h_1^{\perp\gamma}$ is the distribution of linearly polarized photons. 
The effect of $h_1^{\perp\gamma}$ has not be considered in the previous research of TM production. In general case, $h_1^{\perp\gamma}$ is independent of $f_1^{\gamma}$ but there is an upper bound $|h_1^{\perp\gamma}|\leq f_1^{\gamma}$.  However, 
when the photons carry very small longitudinal momenta $k^+$, nearly all the photons are linearly polarized, 
in this case one has $h_1^{\perp\gamma}(x,\vect{k}_T^2)\simeq f_1^{\gamma}(x, \vect{k}_T^2)$~\cite{Li:2019yzy}. This relation will be utilized in the numerical discussions.

The function $H$ can be expressed as
\begin{align}
H_{\mu\rho;\nu\sigma}(p_a,p_b;q)=M_{\mu\rho}(p_a,p_b;q)M^*_{\nu\sigma}(p_a,p_b;q),
\end{align}
where $M_{\mu\rho}$ is the amplitude of $\gamma\gamma\to \rm{pTM}$. It can be calculated as
\begin{align}
	i M^{\mu\rho}\nonumber =&\sqrt{\frac{1}{4\pi}}R(0) \sqrt{\frac{1}{8m^3} } \operatorname{Tr} \bigg\{\bigg(\frac{\slashed q}{2}-m\bigg)\gamma_5 \bigg(\frac{\slashed q}{2}+m\bigg)\nonumber\\
	&~~~~\times\bigg[(-i e\gamma^{\mu})\frac{i}{\frac{\slashed q}{2}-\slashed p_a-m}(-i e\gamma^{\rho}) \nonumber\\
	&~~~~ + (-i e \gamma^{\rho})\frac{i}{\frac{\slashed q}{2}-\slashed p_b-m} (-i e\gamma^{\mu})\bigg]\bigg\}, 
\end{align}
where $m$ is the mass of muon, $R(0)$ is radial wave function of pTM at the origin. The mass of pTM is approximated by $2m$.
Simplify the expression, the amplitude becomes
\begin{align}
i M^{\mu\rho}
=&- \sqrt{\frac{1}{4\pi}}R(0) \frac{2\sqrt{2} e^2}{\sqrt{m}}    \epsilon_T^{\mu\rho}, 
\end{align}
where we have made the approximation $p_a\sim x_a P_A^+ n$ and $p_b\sim x_b P_B^- \bar n$ in the partonic amplitude, {$P_A$ and $P_B$ are the momenta of nucleus $A$ and $B$ (per nucleon), respectively. The collision energy squared $s$ (per nucleon pair) can be expressed as $s=(P_A+P_B)^2\approx 2 P_A^+ P_B^-$}. The transverse Levi-Civita tensor is $\epsilon_T^{\mu\rho}=\epsilon^{\mu\rho\nu\sigma}n_{\nu}\bar n_{\sigma}$.
Then the differential cross section becomes,
\begin{align}
	&\frac{d\sigma}{dy dq_T^2}=\frac{2\pi^2}{s} \alpha^5 \zeta(3) \int   d^2 \vect{p}_{aT}d^2 \vect{p}_{bT}   \delta^2 (\vect{p}_{aT}+\vect{p}_{bT}-\vect{q}_T) \nonumber\\
	& ~~~~~~ \times  \bigg[   f_{1}^{\gamma}(x_a, \vect{p}_{aT}^2) f_{1}^{\gamma} (x_b, \vect{p}_{bT}^2) \nonumber\\
	& ~~~~~~  - h_1^{\perp\gamma} (x_a, \vect{p}_{aT}^2) h_1^{\perp\gamma} (x_b, \vect{p}_{bT}^2) \bigg(\frac{2 \vect{p}_{aT}\cdot \vect{p}_{bT}^2}{\vect{p}_{aT}^2 \vect{p}_{bT}^2} - 1 \bigg) \bigg] ,
\end{align}
where $\alpha\approx 1/137$ is the fine structure constant,  $x_{a,b}=\frac{2m}{\sqrt{s}}e^{\pm y}$, and {$y\equiv \frac12 \ln \frac{q^+}{q^-}$ is the rapidity of pTM, with $q^+$ and $q^-$ denoting the ``$+$'' and ``$-$'' components of the pTM momentum $q$ in the light cone coordinate system, respectively.} A very similar formula for $^1S_0$ quarkonium production has been derived before, which is expressed in terms of the gluon distributions~\cite{Boer:2012bt,Ma:2012hh}. 
When deriving the above results we have adopted 
$	R(0)^2 =\frac{\alpha^3 m^3}{2  n^3}$ with $n$ being the principal quantum number; summing over $n$ leads to a factor $\zeta(3)$. It is interesting to note that the formula doesn't depend on the muon mass $m$ at first glance because all the 
$m$ dependence is through the variables $x_{a,b}=\frac{2m}{\sqrt{s}}e^{\pm y}$. Integrating over $q_{T}$, one can find that the term associated with $h_1^{\perp\gamma}h_1^{\perp\gamma}$ disappears and get
\begin{align}
&	\frac{d\sigma}{dy }
	=\frac{2\pi^2}{s} \alpha^5 \zeta(3) \nonumber\\
&~~~~~~\times\int   d^2 \vect{p}_{aT}d^2 \vect{p}_{bT}        f_{1}^{\gamma}(x_a, \vect{p}_{aT}^2) f_{1}^{\gamma} (x_b, \vect{p}_{bT}^2)      ,
\end{align}
so the linearly polarized photon distribution will not modify the $q_T$ integrated cross section, as well as the total cross section. However, the linearly polarized photon will modify the transverse momentum distribution of the pTM, as we shall demonstrate below. The total cross section can be derived by integrating out $y$,
\begin{align}
	\sigma =  \frac{2\pi^2}{s} \alpha^5 \zeta(3)      \int dy  f_{1}^{\gamma}\left(\frac{2m}{\sqrt{s}}e^{y} \right) f_{1}^{\gamma} \left(\frac{2m}{\sqrt{s}}e^{-y}\right)  ,
\end{align}
where 
$  f_{1}^{\gamma}(x_i)=\int d^2 p_{i T}	  f_{1}^{\gamma}(x_i, \vect{p}_{iT}^2)$ are the integrated photon distributions.

\begin{figure}[!htbp]
	\centering
	\includegraphics[width=0.9\linewidth]{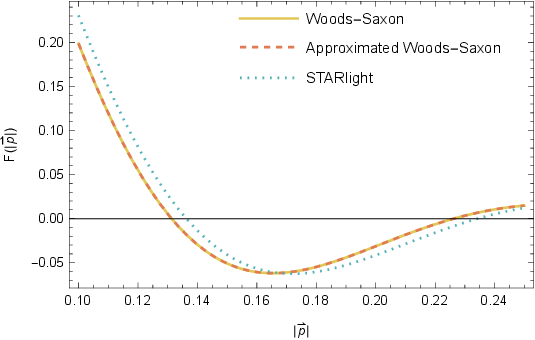}
	\caption{The Woods-Saxon form factor, the form factor adopted by this work, and the form factor in STARlight MC generator of Pb, where $R_{\mathrm{WS}}=6.62$ fm, and $d=0.546$ fm. Too see the differences between the form factors more clearly, we ``zoom in'' by plotting in a particular region. }
	\label{fig:ff}
\end{figure}

According to the Weizs\"{a}cker-Williams method, the unpolarized photon distribution for a nucleus is
\begin{align}
	x f_1^{\gamma}(x, \vect{p}_T^2)=\frac{Z^2 \alpha}{\pi^2} \frac{\vect{p}_T^2}{(\vect{p}_T^2+x^2 M_p^2)^2} F^2 (\vect{p}_T^2+x^2 M_p^2) ,
\end{align}
where $M_p$ is the proton mass, $Z$ is the atom number and $F$ is the electric form factor of the nucleus. The form factor is often parametrized using the Woods-Saxon distribution~\cite{Woods:1954zz}
\begin{align}
	F(|\vec{p}|)=\int d^3 r e^{i\vec{p}\cdot \vec{r}} \frac{\rho^0}{1+e^{\frac{r-R_{\mathrm{WS}}}{d}}} , \label{eq:woodssaxon}
\end{align}
where $\rho_0$ is the normalization factor,  $R_{\mathrm{WS}}$  is the radius, and $d$ is the skin depth.  The exact analytical expression of the form factor $F$ can be derived by doing the Fourier integral.  The result reads
\begin{align}
	 &F(|\vec{p}|) 
	=\frac{4\pi d \rho_0}{|\vec{p}|}\bigg[\pi\operatorname{csch}( \pi d |\vec{p}|)\nonumber\\
	&\times\bigg( \pi  d \sin(R_{\mathrm{WS}} |\vec{p}|)\coth(\pi d |\vec{p}|)-R_{\mathrm{WS}}\cos(R_{\mathrm{WS}} |\vec{p}| )\bigg) \nonumber\\
	&~~~~~~~~~~ - d \operatorname{Im}\Phi(-e^{-\frac{R_{\mathrm{WS}}}{d}}, 2, -i d |\vec{p}|)\bigg], 
\end{align}
where $\Phi(z,s,a)\equiv \sum_{n=0}^{\infty} \frac{z^n}{(n+a)^s}$ is the Hurwitz-Lerch transcendent. The last line in the above expression is extremely tiny because of the exponential, thus one can neglect it and get an approximation of the Woods-Saxon distribution,
\begin{align}
	&F(|\vec{p}|) 
\simeq\frac{4\pi^2 d \rho_0}{|\vec{p}|} \operatorname{csch}( \pi d |\vec{p}|) \nonumber\\
\times&\bigg( \pi  d \coth(\pi d |\vec{p}|) \sin(R_{\mathrm{WS}} |\vec{p}|)-R_{\mathrm{WS}}\cos(R_{\mathrm{WS}} |\vec{p}| )\bigg),  \label{eq:formfactor}
\end{align}
and the normalization factor $\rho_0$ is evaluated as 
\begin{align}
 \rho_0=\frac{3}{4\pi}\frac{1}{R_{\mathrm{WS}}(d^2 \pi^2+R_{\mathrm{WS}}^2)}.
 \end{align}
Numerical result shows that Eq.~\eqref{eq:formfactor} is a very good approximation of Eq.~\eqref{eq:woodssaxon}, and we will adopt it for further convenience in most of the  numerical calculations. We note that the same expression has been derived in Ref.~\cite{Sengul:2015ira}. 
Another commonly used  form factor in literature is the one from the STARlight Monte Carlo (MC) generator~\cite{Klein:1999qj,Klein:2016yzr},
\begin{align}
	F(|\vec{p}|)=&\frac{4\pi\rho^0}{ A |\vec{p}|^3 } \bigg[\sin (R_A |\vec{p}|) - R_A |\vec{p}|  \cos (R_A |\vec{p}| )\bigg] \nonumber\\
	&\times \frac{1}{1+a^2 \vec{p}^2}, \label{eq:starlight}
\end{align}
where $R_A=1.1 A^{1/3}$ fm, $A$ is the mass number of nuclei and $a=0.7$ fm. The normalization factor $\rho^0=\frac{3}{4\pi}\frac{A}{R_A^3}$ is the nuclear density. It is also a good approximation of the Woods-Saxon distribution. In Fig.~\ref{fig:ff}, we plot the Woods-Saxon distribution, the form factor in Eq.~\eqref{eq:formfactor}, and the STARlight form factor of Pb, where $R_{\mathrm{WS}}=6.62$ fm, and $d=0.546$ fm.
The comparison of the total cross sections computed from the two form factors will also be presented later.

Based on the above results, we present the numerical analysis herein, for the Au-Au collisions at the RHIC and Pb-Pb collisions at the LHC.
 The nucleus radii are $R_{\mathrm{WS}}=6.38$ fm for ${}^{197}\mathrm{Au}$ and $R_{\mathrm{WS}}=6.62$ fm for ${}^{207}\mathrm{Pb}$, while the corresponding skin depths are $d=0.535$ fm for ${}^{197}\mathrm{Au}$ and $d=0.546$ fm for ${}^{207}\mathrm{Pb}$~\cite{DeVries:1987atn}. As discussed earlier, one has $h_1^{\perp\gamma}(x,\vect{k}_T^2)\simeq f_1 (x,\vect{k}_T^2)$, i.e., all photons are linearly polarized. The case that photons are partially polarized will be discussed later.

\begin{figure}[!htbp]
	\centering
	\includegraphics[width=0.9\linewidth]{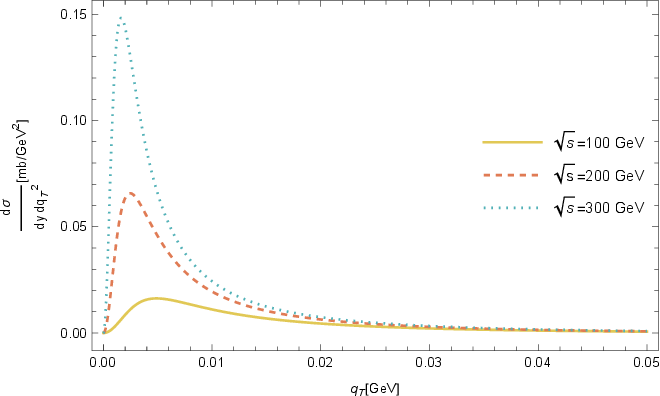}
	\caption{The differential cross section as a function of the pTM transverse momentum $q_T$ in Au Au collisions. We take the collision energy per nucleon pair as $\sqrt{s}=100~\mathrm{GeV}, 200~\mathrm{GeV}, 300$ GeV that are near the RHIC energy. The differential cross section exhibits an increase with the rising values of $\sqrt{s}$.}
	\label{fig:dsigmadydqt2}
\end{figure}

\begin{figure}[!htbp]
	\centering
	\includegraphics[width=0.9\linewidth]{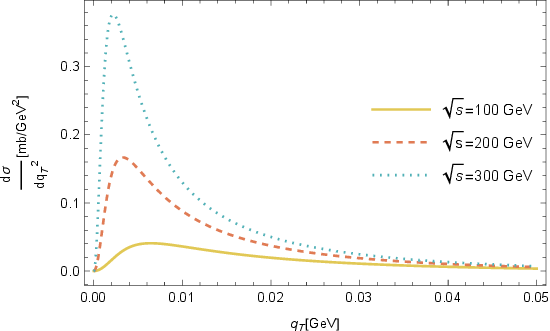}
	\caption{The $y$-integrated differential cross section as a function of $q_T$ with the collisions energy $\sqrt{s}=100~\mathrm{GeV}, 200~\mathrm{GeV}, 300$ GeV per nucleon pair in Au-Au collisions, where $y$ is the rapidity of pTM. }
	\label{fig:dsigmadqt2}
\end{figure}

To delve into the differential cross section dependence on $\sqrt{s}$, we fix $y=0$. The resulting plot illustrates the differential cross section versus transverse momentum $q_{T}$ of pTM. We consider three distinct center-of-mass energies per nucleon pair: $\sqrt{s}=100$ GeV, $200$ GeV, and $300$ GeV of Au-Au collisions, which are around the RHIC collision energy $\sqrt{s}=200$ GeV per nucleon pair. The graphical representation is depicted in Fig.~\ref{fig:dsigmadydqt2}. Integrating over the rapidity $y$ one can have the $y$-integrated differential cross section $d\sigma/d q_T^2$, which is shown in Fig.~\ref{fig:dsigmadqt2}. One can find that the differential cross section increases with the rising of $\sqrt{s}$, which is because that the photon density increases when $x$ goes smaller. 

\begin{figure}[!htbp]
	\centering
	\includegraphics[width=0.88\linewidth]{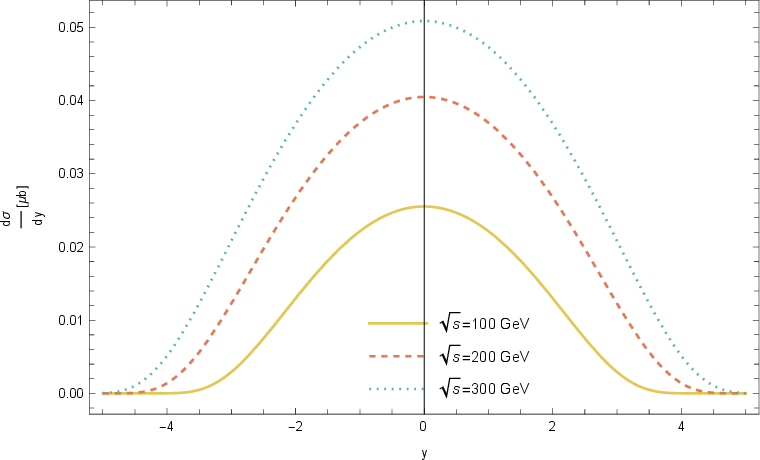}
	\caption{The $q_T$ integrated differential cross section $d\sigma/dy$ as a function of $y$ in Au-Au collisions, where $y$ denotes the rapidity of pTM, and the collision energy $\sqrt{s}=100~\mathrm{GeV}, 200~\mathrm{GeV}, 300$ GeV per nucleon pair in the center-of-mass frame.}
	\label{fig:dsigmadyrhic}
\end{figure}

\begin{figure}[!htbp]
	\centering
	\includegraphics[width=0.9\linewidth]{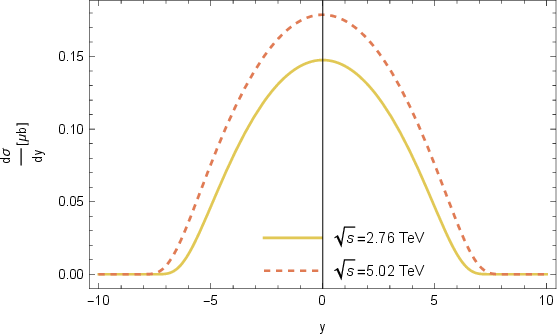}
	\caption{The $q_T$ integrated differential cross section $d\sigma/dy$ as a function of $y$ in Pb-Pb collisions, where $y$ denotes the rapidity of pTM. We take the collision energy $\sqrt{s}=2.76$ TeV and $5.02$ TeV per nucleon pair in the center-of-mass frame. }
	\label{fig:dsigmadylhc}
\end{figure}

To see how the produced pTM is distributed with its rapidity, we 
integrate over $q_T$ and get the $q_T$-integrated differential cross section $d\sigma/dy$ as a function of $y$---the rapidity of pTM---which is presented in Fig.~\ref{fig:dsigmadyrhic} for Au-Au collisions and Fig.~\ref{fig:dsigmadylhc} for the Pb-Pb collisions. Once more, we let $\sqrt{s}=100~\mathrm{GeV}, 200~\mathrm{GeV}, 300$ GeV for Au-Au mode, and take $\sqrt{s}=2.76~\mathrm{TeV}, 5.02$ TeV for Pb-Pb mode which corresponds to the center-of-mass energy per nucleon pair at the LHC. One can find that the rapidity of pTM at the RHIC is around $|y|\lesssim 5$, while $|y|\lesssim 8$ at the LHC. 

\begin{figure}[!htbp]
	\centering
	\includegraphics[width=0.9\linewidth]{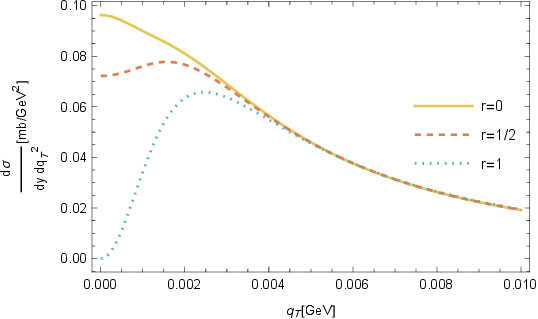}
	\caption{The differential cross section computed with $r=0$, $r=1/2$ and $r=1$, in Au-Au collisions, where $r$ is the degree of photon polarization, and $y$ is the rapidity of pTM. The collision energy per nucleon pair is taken as $\sqrt{s}=200$ GeV. }
	\label{fig:degreeofpolarization}
\end{figure}

To see how the polarization of photon affects the production of pTM, we introduce the degree of polarization $r$, which is defined as the ratio of the linearly polarized and unpolarized photon distributions, i.e.,  
$	r\equiv  {h_1^{\perp\gamma} (x, \vect{k}_T^2)}/{f_1^{\gamma}(x, \vect{k}_T^2)} $.  $r$ is generally a function, but for the sake of simplicity, we assume $r$ is a number and present the differential cross sections with $r=0, 1/2, 1$ in Fig.~\ref{fig:degreeofpolarization}. When $r=0$, there is no linearly polarized photon, while for $r=1$ all of the photons are linearly polarized.  One can find that the linearly polarized photons strongly effect the transverse momentum dependence in the small-$q_T$ region, while for larger $q_{T}$ the differential cross section is not affected.  It is interesting to figure out that, the differential cross section for fully linearly polarized photons decreases to zero when $q_T\to 0$, and the differential cross section approaches the maximum value around several MeV.  

\begin{figure}
	\centering
	\includegraphics[width=0.9\linewidth]{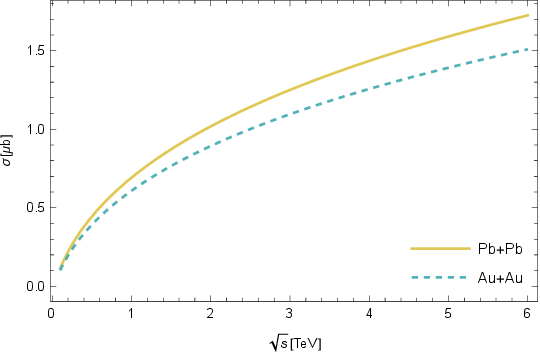}
	\caption{The total cross sections of pTM production in Au-Au and Pb-Pb collisions.}
	\label{fig:totalcrossection}
\end{figure}

Integrating  out $y$ and $q_{T}$ leads to the total cross section.  We plot the total cross sections as the functions of $\sqrt{s}$ for both Au-Au and Pb-Pb collisions, which is shown in  Fig.~\ref{fig:totalcrossection}. Again, one can conclude that the total cross section increases when $\sqrt{s}$ increases.  We list the total cross sections  in Table~\ref{table:totalcrosssection}. The cross sections of pTM production reported in~\cite{Ginzburg:1998df} are 0.15~$\mu$b for gold-gold mode at RHIC, and 1.35~$\mu$b for lead-lead collisions at LHC. Our results in Table~\ref{table:totalcrosssection} are larger, but in agreement with their results in magnitude.  As a coproduct of this analysis, the cross sections for the production of para-positronium (pPM) and true para-tauonium (pTT) are also presented in Table~\ref{table:totalcrosssection}. We note that the total cross section for true tauonium production via two photon fusions in UPCs was calculated in Refs.~\cite{dEnterria:2022ysg,Shao:2022cly,dEnterria:2023yao}; our result is larger because the UPC events are not singled out in the traditional TMD formalism.
\begin{table*}[!ht]
	\centering
	\begin{tabular}{c|c|c|c}
		\hline\hline
	$\sqrt{s}$ (TeV)	& 0.2 (Au-Au)  & 2.76 (Pb-Pb)&  5.02 (Pb-Pb)  \\
		\hline
		pTM cross section ($\mu$b)& 	0.20 & 1.20  & 1.59  \\
		\hline
		pPM cross section ($\mu$b)& 	$1.36\times 10^5$ & $5.85\times 10^5$  & $4.01\times 10^6$  \\
		\hline
		pTT cross section ($\mu$b)& 	$9.68\times 10^{-6}$ & $6.55\times 10^{-4}$ & $1.08\times 10^{-3}$ \\
		\hline\hline
	\end{tabular}
\caption{The total cross sections of pTM production in two gamma fusion in Au-Au and Pb-Pb collisions. In the second row we list the cross sections of para-positronium production; while in the third row we list the cross sections for true para-tauonium.  }\label{table:totalcrosssection}
\end{table*}

\begin{figure}
	\centering
	\includegraphics[width=0.9\linewidth]{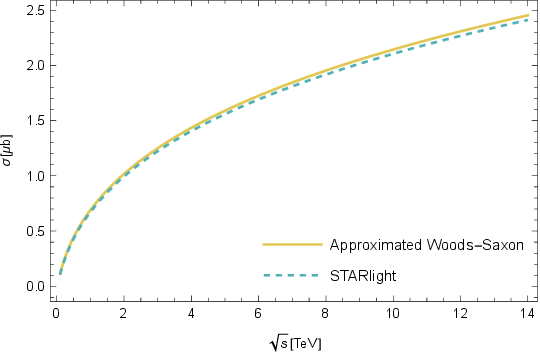}
	\caption{The total cross section of pTM production in Pb-Pb collisions, computed with the form factor in Eq.~\ref{eq:formfactor} and the one from STARlight MC generator. }
	\label{fig:sigmawsandsl}
\end{figure}

In Fig.~\ref{fig:sigmawsandsl}, we explore the effects of different choices of nuclei form factors on the total cross sections. It indicates that the result calculated with the  form factor in Eq.~\eqref{eq:formfactor} matches well with the result calculated with the form factor adopted by STARlight, especially when $\sqrt{s}$ is not too large; when $\sqrt{s}$ is large, there may be visible differences between the two cross sections. 

\begin{figure}[!htbp]
	\centering
	\includegraphics[width=0.95\linewidth]{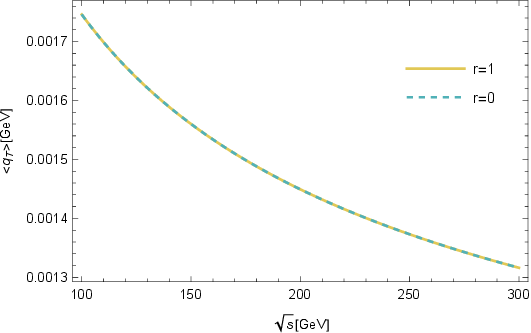}
	\caption{The average transverse momentum $\langle q_{T} \rangle$ as the function of $\sqrt{s}$ in Au-Au collisions, computed in the case of $r=1$ and $r=0$, where $r$ is the degree of polarization.}
	\label{fig:qt}
\end{figure}

It is also intriguing to explore the average transverse momentum of pTM. We define the average transverse momentum as
\begin{align}
	\langle q_{T} \rangle \equiv \frac{\int dy dq_T^2 q_T \frac{d\sigma}{dy d q_T^2}}{\int dy dq_T^2  \frac{d\sigma}{dy d q_T^2}}. 
\end{align}
With the differential cross sections calculated above, one can compute the numerical results of $\langle q_{T} \rangle$, which is plotted 
in Fig.~\ref{fig:qt}. We take the Au-Au collisions as an example. At the RHIC energy $\sqrt{s}=200$ GeV,  $	\langle q_{T} \rangle\simeq 1.4 $ MeV. One can find that the average transverse momentum decreases when $\sqrt{s}$ increases. We plot $\langle q_{T} \rangle$ for both $r=0$ and $r=1$ cases, and find that the two curves coincides. Therefore, although the linearly polarized photons modify the shape of $q_T$ distribution, the average value $\langle q_{T} \rangle$ is not affected by the polarization. 

To summarize, we study the true para-muonium production with photon-photon fusions in heavy ion collisions. The transverse distribution of true para-muonium is taken into explored, and to achieve this purpose after taking the linearly polarized photon distribution into account. Our results indicate that the differential cross section of true para-muonium production in the small $q_T$ region is affected by the linearly polarized photons in nuclei significantly, and exhibits a maximum when $q_T$ is around several to 10 MeV at the {RHIC, while the maximum of differential cross section is located at lower $q_T$ at the LHC}. {Detecting pTMs with such low momenta presents a challenge in current experiments at the LHC, as the minimum transverse momentum measured is approximately 1 GeV~\cite{ATLAS:2008xda,CMS:2008xjf,LHCb:2008vvz,ALICE:2008ngc}. However, it is crucial to acknowledge that transverse momentum distributions have the potential to serve as valuable guides for future experimental analyses.}
The total cross section is also computed, indicating that significantly considerable amounts of true para-muonium should be produced. 
Our work can also be extend to the true ortho-muonium production. 

\section*{Acknowledgments}

J. P. D. thanks H. B. Li and B. S. Zou for useful discussions. S. Z. acknowledges A. Arbuzov, S. J. Brodsky, R. Lebed, and W. Wang for communications and discussions on related topics. We also thank D.~d'Enterria and H.~S.~Shao for valuable comments. The work of J. P. D. is supported by the National Natural Science Foundation of China (Grants No. 12165022) and Yunnan Fundamental Research Project under Contract No. 202301AT070162.

\end{document}